\documentclass[]{elsart}

\usepackage{times}
\usepackage{epsf}
\usepackage[numbers,sort&compress]{natbib}
\usepackage{amsmath}
\usepackage{amsfonts}
\usepackage{amssymb}
\usepackage{bm}
\usepackage{bbm}
\usepackage{graphicx}
\usepackage{epsfig}
\usepackage{latexsym}
\usepackage{nicefrac}

\usepackage{dcolumn}

\newcommand{\beq}{\begin{equation}}
\newcommand{\eeq}{\end{equation}}
\newcommand{\bea}{\begin{eqnarray}}
\newcommand{\eea}{\end{eqnarray}}
\newcommand{\hf} {\frac{1}{2}}

\newcommand\Ref[1]     {Ref.\,\cite{#1}}

\newcommand\eqn[1]     {Eq.\,(\ref{#1})}

\newcommand\fig[1]     {Fig.\,{\ref{#1}}}

\def\tu{{\tilde u}}

\def\ord#1{{\cal O}(#1)}

\def\mr#1{{\mathrm{#1}}}

\def\eq#1{(\ref{#1})}

\begin{document}

\begin{frontmatter}

\title{Renormalizable parameters of the sine-Gordon model}

\author{S. Nagy$^1$, I. N\'andori$^{2}$, J. Polonyi$^3$, K. Sailer$^1$}

\address{$^1$Department of Theoretical Physics, University of Debrecen,
Debrecen, Hungary\\
$^2$Institute of Nuclear Research of the Hungarian Academy of Sciences,
H-4001 Debrecen, P.O.Box 51, Hungary\\
$^3$ Institute for Theoretical Physics, Louis Pasteur University,
Strasbourg, France}
\date{\today}

\begin{abstract}
The well-known phase structure of the two-dimensional sine-Gordon model is 
reconstructed by means of its renormalization group flow, the study of the
sensitivity of the dynamics on microscopic parameters. Such an
analysis resolves the apparent contradiction between the phase structure and the
triviality of the effective potential in either phases, provides a
case where usual classification of operators based on the linearization of 
the scaling relation around a fixed point is not available and 
shows that the Maxwell-cut generates an unusually strong universality at
long distances. Possible analogies with four-dimensional Yang-Mills 
theories are mentioned, too.
\end{abstract}


\end{frontmatter}

{\it Introduction} The phase structure and renormalizability of field theoretical
models are well understood in the context of the renormalization group:
Phase transitions belong to the singularities of the effective IR coupling 
constants as the functions of the UV parameters and the renormalizable
parameters are the relevant or marginal coupling constants of the UV fixed point.
The two-dimensional sine-Gordon (SG) model has a well-known phase structure
and renormalization group flow but does not easily fit into the general scheme.
Our aim in this work is the clarification of these issues by a careful 
renormalization group study of the SG model.

The SG model, defined by the action
\beq\label{sqcont}
S=\int_x\left[\frac12(\partial_\mu\phi_x)^2+u_1\cos(\beta\phi)\right],
\eeq
has an ionized (massless, strong coupling, non-renormalizable) 
phase for $\beta^2>8\pi$ and a molecular (massive, weak coupling, renormalizable) 
phase for $\beta^2<8\pi$. The molecular phase is perturbatively equivalent with 
the neutral sector of the massive Thirring model \cite{Coleman_bos,sg-mt}
and the neutral Coulomb-gas \cite{Samuel}. Perturbation expansion was used
to conjecture the same universal behavior around $\beta^2=8\pi$ \cite{Amit,Balog}
and phase structure \cite{KTB} as that of the planar X-Y model. Simple
comparison of the lattice regulated SG model with the planar X-Y model
provides a non-perturbative renormalization group flow for the SG model \cite{huang}.
The direct lattice regularization of the SG model makes it evident
that our model, given by Eq. \eq{sqcont}, has aperiodic kinetic energy.
In fact, the first term in the action is periodic for space-time independent,
$p=0$ Fourier mode only. Its aperiodicity for space-time dependent fluctuations
suppresses configurations which correspond to the vortices of the X-Y model.
As a result, the model
has no or few renormalizable coupling constants in the ionized and the
molecular phases, respectively \cite{huang}. It has been argued that the appropriate 
order parameter, provided by the soliton dynamics, is the topological 
susceptibility \cite{nandori}, because 
the symmetry with respect to the fundamental group $\pi_1(U(1))$, 
\beq\label{fundgr}
\phi_x\to\phi_x+\Delta,
\eeq
where
\beq\label{perl}
\Delta=\frac{2\pi}{\beta}
\eeq
is the period length of the local potential, is broken dynamically in the molecular phase.

But some complication arises by the similarity of the deep infrared IR
dynamics in the two phases of the SG model \cite{nandori}. The rather trivial
observation, namely that the only non-singular effective potential which is periodic 
and convex is the constant shows that the Maxwell-construction washes away
the differences of the phases in the deep IR regime. 
How can this be reconciled with the overwhelming
evidences about the two differing phases and the different numbers of the
renormalizable parameters? We show that the difference of the two phases
appears in the effective potential expressed in units of the running cutoff.
This potential produces weak effects in the deep IR but can 
be used to identify the phase structure because it produces singularities in
the sensitivity of the dynamics at a given observational scale as the
function of the bare parameters, given at the cutoff scale \cite{janosRG}. 
Such a global use of the renormalized trajectory is needed because of the
non-triviality of the IR scaling laws, the inherent non-linear nature of the
renormalization group trajectory at the IR fixed point of the ionized phase.

{\it Blocking and condensate} The blocking relation for the Euclidean Wilsonian action,
$S_k[\phi]$
to be integrated from the initial condition imposed at the cutoff $k=\Lambda$, is
\beq\label{blrel}
e^{-S_{k-\Delta k}[\phi]}=\int D[\phi']e^{-S_k[\phi+\phi']}
\eeq
where the Fourier amplitudes of $\phi$ and $\phi'$ are restricted to
wave numbers $0\le|p|\le k-\Delta k$, and $k-\Delta k\le|p|\le k$, respectively
\cite{wh}, in the spirit of the differential renormalization group (RG)
\cite{polc,diff}. The functional integral is over modes whose scale belongs to a finite 
interval, therefore the dependence of the blocked action $S_{k-\Delta k}[\phi]$
on the parameters of the original action, $S_k[\phi]$, is regular \cite{regrg},
no phase transition-like singularities \cite{srg,srgh,srgnu} are allowed.
The loop-expansion of the functional integral gives
\beq\label{eveqwsp}
S_{k-\Delta k}[\phi]=S_k[\phi+\phi'_0]+\hf\mr{Tr}'\ln\frac{\delta^2S_k[\phi+\phi'_0]}
{\delta\phi}{\delta\phi}+\ord{(\frac{\Delta k}{k})^2},
\eeq
where $\phi'_0$ is the background field dependent saddle point \cite{janostree}. 

The local potential approximation (LPA) for the SG model consists of the 
projection of the evolution equation into the functional space 
\beq
S_k = \int_x\left[\frac12(\partial_\mu\phi_x)^2+U_k(\phi_x)\right]
\label{action}
\eeq
where the local potential is given by the Fourier series
\beq
U_k(\phi)= \sum_{n=1}^\infty u_{n}(k)\cos(n\beta\phi).
\label{per}
\eeq
The Wegner-Houghton equation \cite{wh} reads
\beq
\label{WHdim}
\left(2 + k\partial_k \right) {\tilde U}_k ( \phi) = 
- \frac1{4\pi}  \ln\left(1 + {\tilde U}^{(2)}_k (\phi) \right),
\eeq
when the vanishing of the saddle point is assumed, ${\tilde U}_k = k^{-2} U_k$ 
is the dimensionless potential and ${\tilde U}^{(2)}_k (\phi)=\partial^2_\phi{\tilde U}_k (\phi)$.

The saddle point which indicates the presence of a condensate in the 
system, the hallmark of spontaneous symmetry breaking, appears at the scale where
the Euclidean inverse propagator, $\delta^2S_k/\delta\phi\delta\phi$, looses its 
positive definiteness. The negative value of the inverse propagator indicates that the 
action has unstable minimum for vanishing strength of the field components
to be eliminated. The action is bounded from below therefore the true minimum 
is degenerate, implying highly populated modes, a condensate whose strength is 
stabilized by the balance between kinetic and potential energies. 

The loss of positive definiteness of the propagator corresponds to the change of 
sign of the argument of the logarithmic function in Eq. \eq{WHdim} in our approximation.
Similar phenomenon has already been observed in the case of the scalar $\phi^4$ 
model in Ref. \cite{riwe}, too. Once the tree-level contribution to the blocking relation
appeared we neglect the loop contributions and the evolution is governed by the 
tree-level blocking relation $S_{k-\Delta k} [\phi] = \min_{\phi'}(S_k[\phi + \phi'])$.
We search the saddle point among plane waves,
$\phi_x'=\rho_k\cos(kn_k\cdot x+\theta_k)$ with $n_k$ being a unit vector, and the result 
is \cite{janosRG,nandori}
\beq
\label{treedim}
{\tilde U}_{k-\Delta k}(\phi) = \min_{\rho} \left[\rho^2 +\hf
\int_{-1}^{1} du {\tilde U}_{k}(\phi + 2\rho \cos(\pi u)) \right],
\eeq
replacing Eq. \eq{WHdim} in the unstable region. It is easy to show that the RG 
flow in LPA preserves the period length of the potential. 

The saddle point reflects the macroscopic population of a mode and is a well
known phenomenon in theories with spontaneously broken symmetry. For instance,
the symmetry broken vacuum of the simple $\phi^4$ model has a condensate 
in the mode $p=0$. What is new in the saddle point occurring at finite
$k$ in the blocking equation is that it is space-time dependent. There are 
such inhomogeneous, macroscopically populated modes in the $\phi^4$
model if the absolute magnitude of the space-time average of the field 
expectation value is constrained to smaller values than in the symmetry broken 
vacuum. Such a system is in a mixed phase because the constraint
is satisfied by a domain structure where each domain has a stable, clusterising 
vacuum expectation value of the field. The inhomogeneous saddle points represent 
the domain walls. 

An important consequence of the inhomogeneous saddle point is the appearance
of soft modes, the analogies of the Goldstone modes, even if the internal symmetry, 
broken spontaneously, is discrete. In fact, the parameters $\theta_k$ and 
$n_k$ are zero modes arising from the breaking of the continuous space-time 
symmetries, translations and rotation, respectively, by the inhomogeneous saddle point.
They are the soft, non-perturbative modes of the domain structure and 
generate the Maxwell-cut, the degeneracy of the 
true ground state as the function of the field expectation value within 
the mixed phase \cite{janostree}.

Do we expect condensate in the SG model? The stability of the kinks
originate form the spontaneous breakdown of the fundamental group symmetry
\eq{fundgr}. Thus condensate is expected when so long the solitons are stable.
The presence of virtual kink-anti kink polarisation of the vacuum indicates 
that this condensate contains inhomogeneous modes, too.

After having clarified the dynamical role played by inhomogeneous condensates
in our model we return to the evolution equation \eq{eveqwsp}.
The change of the sign of the propagator signals a singularity in the second
term on the right hand side, an IR Landau pole which is always
an artifact of the truncation. Although the condensate modifies the dynamics 
and restores the positive definiteness of the propagator nevertheless the question
remains: What happens with the exact solution to the evolution equation? 
An important test is passed by the numerical integration of the evolution 
equation, it reproduces regular scale dependence in the blocking relations, 
as expected. But the sharp cutoff in the evolution equation \eq{blrel} for the 
Wilsonian action prevents us to go beyond the LPA. One should use 
smooth cutoff which can be implemented for the Wilsonian action \cite{polc} 
or the effective action \cite{diff}. The former would be more appropriate in
our case because the evolution equation for the effective action leads to 
degenerate action \cite{effac} but this is the natural manifestation of the 
convexity of the Legendre transform and completely hides the dynamics behind 
the Maxwell-cut. We do not pursue this direction of improvement in the
present work, instead we assume that the appearance of inhomogeneous saddle 
points in our truncation scheme is already the correct reflection of 
macroscopically populated states, a spontaneously broken symmetry 
with respect to the fundamental group transformation in the vacuum.

The solution of the RG equation \eq{WHdim} in the UV scaling regime, $k^2\gg|U_\Lambda^{(2)}|$, is
\beq
\tilde u_{n}(k) = \tilde u_{n}(\Lambda)
\left(\frac{k}{\Lambda}\right)^{n^2 \frac{\beta^2}{4\pi}-2}
\label{dluev}
\eeq
after ignoring contributions $\ord{|U_\Lambda^{(2)}|^2/k^4}$ in \eqn{WHdim}, displaying the
well-known critical point at $\beta^2=8\pi$. The relevant coupling constants 
of the UV fixed point are called renormalizable because those parameterize the 
dynamics at finite scales when the cutoff is removed. The two-dimensional 
scalar model with polynomial interactions possesses infinitely many renormalizable 
operators. In contrast, the periodicity of the potential of the SG model seems to
make all parameters of the local potential UV irrelevant, i.e. non-renormalizable in 
the ionized phase, $\beta^2>8\pi$ and to allow only a few renormalizable parameters 
in the molecular phase, $\beta^2<8\pi$. But one has to consider the renormalized 
trajectory globally, by taking into account  the IR scaling laws in order to find the
free parameters of the renormalized dynamics. In fact, the number  of these
parameters might be more or less if the IR scaling has new relevant operators 
or stronger universal features, respectively \cite{janosRG}. We now present
a global study of the renormalization group flow to point out that the 
local analysis at a given fixed point is actually not reliable in the SG model.

The solution of the RG equation can 
only be obtained numerically in the IR region because of the mixing of the 
different Fourier modes and the appearance of condensate for $\beta^2<8\pi$ 
\cite{nandori}. It was also shown in \Ref{nandori} that the effective potential of the SG
model is rendered to a field-independent constant by the requirements of 
periodicity in the internal space and convexity. Therefore, all the dimensionful
couplings $u_n(k)$ should tend to zero in the IR limit. Nevertheless according to
the expectation value of the topological susceptibility as the
disorder parameter of the model one can distinguish two phases, an ionized one
for $\beta^2>8\pi$, and a molecular one for $\beta^2<8\pi$.
As we show below this feature is reflected in the different shape of the dimensionless
effective potential in the two phases.

\begin{figure}[ht]
\includegraphics[width=8cm]{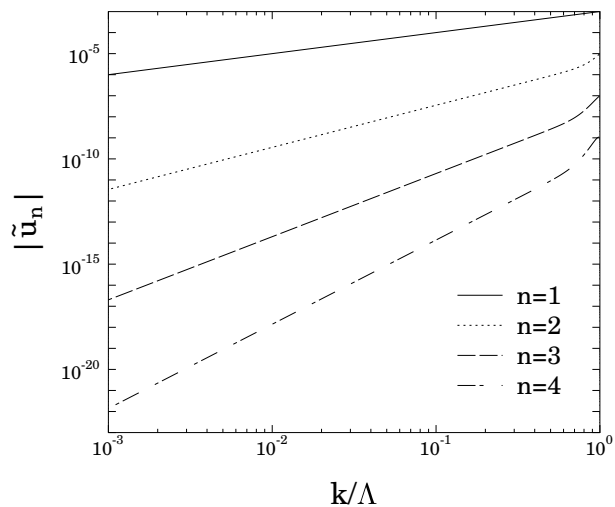}
\caption{The IR scaling law at $\beta^2=12\pi$ for various Fourier amplitudes.
\label{irsg}}
\end{figure}

{\it Ionized phase} There is no spinodal instability and the renormalized trajectory
can be well approximated in the IR scaling regime by the simple power law 
\beq\label{irdlu}
\tu_n = c_n(k/k_0)^{n\eta},
\eeq
with $\eta\ge 0$, $k_0$ being some scale parameter.
In fact, the evolution equation reads as
\beq
(2+n\eta)nc_n k^{n\eta} = \frac{\beta^2}{4\pi}n^3c_nk^{n\eta}+\hf\beta^2
\sum_{s=1}^\infty s A_{n,s}(2+s\eta)c_s k^{s\eta}
\eeq
with $A_{n,s}(k)=(n-s)^2\tu_{|n-s|}-(n+s)^2\tu_{n+s}$,
when this assumption is made. For $n=1$ one finds $\eta =\beta^2/4\pi-2>0$,
$c_1=\tilde u_1(\Lambda)(k_0/\Lambda)^\eta$ and the cases $n>1$ lead to the
 recursion relation
\beq
c_n=\frac{\hf\beta^2 \sum_{s=1}^{n-1}  (2+s\eta)s(n-s)^2c_{n-s}c_s}
{n(2+n\eta-n^2\frac{\beta^2}{4\pi})},
\label{c_n}
\eeq
expressing $c_n$ in terms of $c_1$, $c_n=(-1)^{n+1}\tilde u_1^n(\Lambda)R_n$, 
where $R_1=1$ and the $R_n$ satisfying the recursion relation \eq{c_n}
becomes independent of the bare couplings. This scaling law is confirmed
by the numerical results, shown in \fig{irsg} for $\eta=1$, i.e. $\beta^2=12\pi$.
Further numerical support is that the ratio $R_n= |\tu_n(k)|/\tu_1^n(k)=|c_n|/(c_1)^n$ 
becomes $k$-independent and the local potential becomes independent of
$u_n(\Lambda)$, $n>1$ in the IR region.

A remarkable complication  takes place at the IR fixed point, $\tilde u^\star_n=0$.
The upper harmonics with $n>1$ decrease too fast as $k\to0$, 
$\ord{\tilde u_n}=\ord{\tilde u_1^n}$, and the renormalizable trajectory
is not linearizable around the fixed point. The problem is that one 
tacitly assumes in the usual argument in linearizing the blocking relation
$\tilde u'_n=B_n(\tilde u)$,
\beq
\tilde u_n'-\tilde u_n^\star = 
(\tilde u_m'-\tilde u_m^\star)\partial_m B_n(\tilde u^\star)+
\ord{(\tilde u_m'-\tilde u_m^\star)^2},
\eeq
that the deviation of each coupling constant from the fixed point values is of the same
order of magnitude. As a result, we do not have the usual classification of
operators and a more complete study, based on the global features of the
renormalization group trajectory is needed to determine the number of free
parameters of the renormalized dynamics. The results like the one shown in Fig.
\ref{irsg} indicate that scaling laws characterized by critical exponents
are actually recovered in a non-trivial, non-linear manner with no relevant 
parameter. This result is in agreement with the picture suggested
by the analogy with the X-Y model \cite{huang} where the vortex fugacity is
found to be the only relevant parameter in this phase. But these configurations
have divergent action for the SG model defined in terms of non-periodic
space-time derivatives. In other words, our results correspond to the vanishing
fugacity hyperplane in the space of coupling constants. 

\begin{figure}[ht]
\includegraphics[width=8cm]{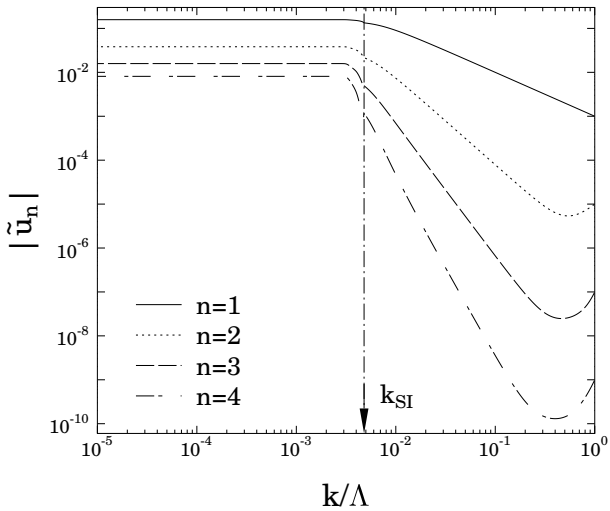}
\caption{
Scale-dependence of the couplings $\tilde u_n$, with $n=1\ldots4$ at
$\beta^2=4\pi$. The scale $k_\mr{SI}$ where the spinodal instability appears is
shown explicitly.
\label{sgspin}}
\end{figure}

{\it Molecular phase} The renormalized trajectory follows scaling laws similar to those of the 
ionized phase at the beginning. Namely, the asymptotic scaling law \eq{dluev} 
in the UV regime ends at a crossover beyond which the scale dependence 
$\tu_n\sim k^{n\eta}$ is encountered. The crucial difference between
the two phases appears in this region in what this scaling law does
not extend to $k=0$ as in the ionized phase, rather it is interrupted 
by a further crossover. The period length of the potential is larger than in the
ionized phase and as a result the symmetry $\phi_x\to\phi_x+2\pi/\beta$,
belonging to the fundamental group of the theory, is broken spontaneously 
at low energy where the potential has more chance to localize the field around
a given minimum by the formation of inhomogeneous saddle-points to the blocking relation. 
The second crossover mentioned above, a spinodal instability appears when the 
propagator diverges, $k_\mr{SI}^2+U^{(2)}_{k_\mr{SI}}(\phi)=0$. 
The blocked action can easily be determined when $\beta^2\to8\pi$ from below 
because $\tu_1$ is the only renormalizable coupling constant.
Due to $k_\mr{SI}/\Lambda\ll 1$ the UV irrelevant couplings die out
and it is sufficient to keep track of $\tu_1$ only, i.e.
one can estimate the value of $k_\mr{SI}$ by calculating the
potential $U_{k}(\phi)$ as if it would contain a single Fourier mode only.
Using $\tu_1(k)=\tu_1(\Lambda) (k/\Lambda)^\eta$ we find
\beq
k^2_\mr{SI} =
\Lambda^2\left(\beta^2 \tu_1(\Lambda)\right)^{\frac{8\pi}{8\pi-\beta^2}}.
\label{kc}
\eeq
We used here the fact that the minimum of $U^{\prime\prime}_k(\phi)$ lies at 
$\phi=0$ in the case of a single coupling. The numerically determined running of 
the coupling constant $\tu_1(k)$ is depicted in Fig. \ref{sgspin}. 
Note that the length scale $1/k_\mr{SI}$ where the spinodal instability
occurs is not an analytic function of $\beta$ at $\beta^2=8\pi$. 
This situation resembles to the non-analyticity of the correlation length 
at the Kosterlitz-Thouless phase transition point \cite{KTB}.

\begin{figure}[ht]
\includegraphics[width=8cm]{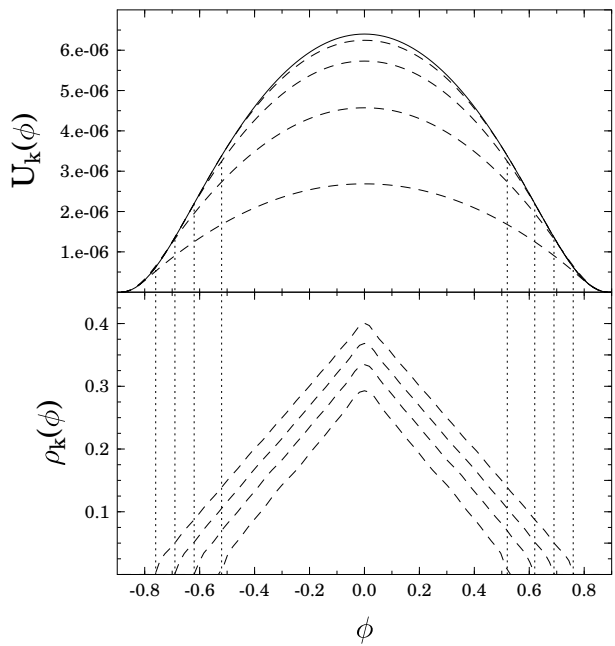}
\caption{The dimensionful potential and the amplitude of the
condensate are plotted for decreasing values of the cut-off for $\beta^2=4\pi$
as the functions of the homogeneous background field. 
The potential shown by solid line 
corresponds to $k=k_\mr{SI}\approx 0.0046$ when the spinodal instability
appears. The potential flattens out and the condensate grows as $k$ is decreased.
\label{pot}}
\end{figure}

Once the cutoff has reached the spinodal instability region its further decrease 
strengthens the condensate and flattens the potential
in order to arrive at a constant potential for $k\to 0$, being the only
function which is simultaneously convex and periodic \cite{nandori}.
The numerical results, displaying this phenomenon are presented in Fig. \ref{pot}.

Though the dimensionful effective potential is flat in both phases,
the dimensionless effective potential $\tilde U_0(\phi)$
is not flat in the molecular phase where one has
$\tilde U_0(\phi)=-\hf(\phi-n\Delta)^2$ for $(n-\hf )\Delta\le\phi<(n+\hf)\Delta$,
with $n=0,1,\ldots$ and $\Delta$ being given by Eq. \eq{perl}.
Both the linear dependence of the strength of the saddle point on the background
field and the quadratic shape of the dimensionless potential can be
understood by simple analytical considerations \cite{pjhep}.

Note that the spinodal instability and condensation also induce an effective potential of
parabolic shape when the periodic bare potential is replaced by a polynomial one. 
The only difference is the lack of periodicity in the latter case. The universal parabolic shape is a direct
consequence of the LPA and the Maxwell-cut which have to be performed on the
unstable region of the effective potential \cite{janostree,gribov}.

\begin{figure}[ht]
\includegraphics[width=8cm]{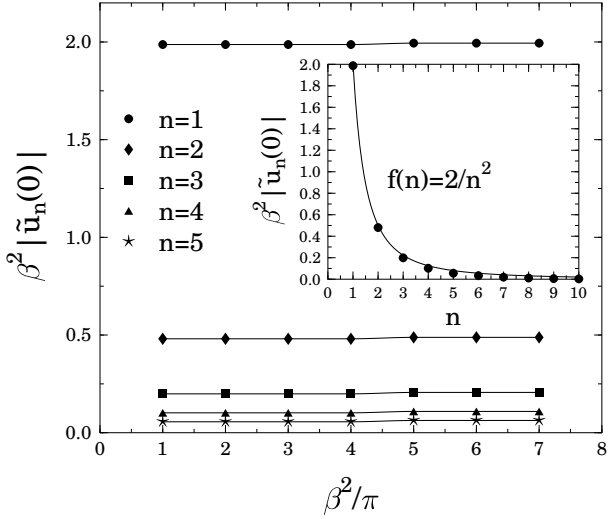}
\caption{
The magnitudes of the first few 
dimensionless couplings are plotted as the function of $\beta$.
The inset shows that the $\beta$-dependence can be easily factorized
and the index-dependence also obtained.
\label{para}}
\end{figure}

In order to make the approach of the potential to a parabola more apparent 
we numerically determined $\beta^2\tu_n(0)$ and found that it is a universal 
function of $n$, $\beta^2\tu_n(0)=2(-1)^n/n^2$ for $\tu_1(\Lambda)>0$, 
cf the inset of Fig. \ref{para}. The corresponding dimensionless
effective potential is
\beq
\tilde U_{k\to 0}(\phi) =
\frac2{\beta^2}\sum_{n=1}^\infty (-1)^{n+1}\frac{\cos(n\beta\phi)}{n^2}=
-\hf\phi^2,  ~~~~\phi\in [-\pi/\beta,\pi/\beta]
\label{effpotmol}
\eeq
apart from a field independent constant. This parabola is repeated periodically
along the $\phi$ axis. It is apparent that such a periodic
potential is the solution of \eqn{treedim}. 

\begin{figure}[ht]
\includegraphics[width=8cm]{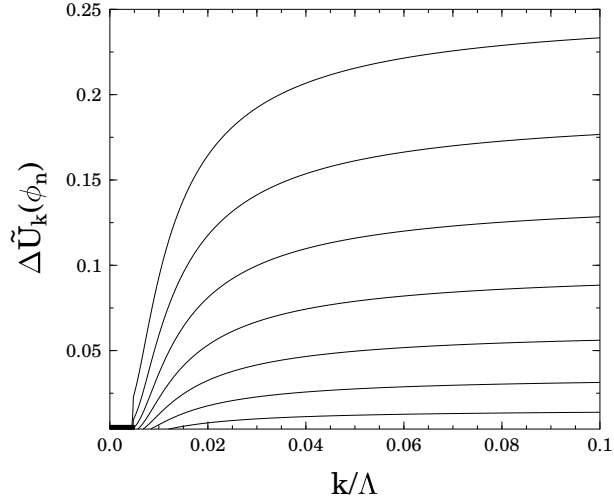}
\caption{
The deviation of the dimensionless blocked potential from
the parabolic shape at different values, $\phi_n=n\pi/(10\beta)$ with $n=0,1,\ldots,10$,
of the field variable. The deviation increases with increasing $n$.
\label{devpar}}
\end{figure}

How is the parabolic shape formed as we approach the IR end point?
The exactly parabolic shape part of the dimensionless blocked potential 
appears at $k=k_\mr{SI}$ only and spreads over larger field region as
the cutoff is further lowered. But the potential approaches a parabola-looking
shape already before the appearance of the condensate as demonstrated in 
\fig{devpar}, where the difference 
$\Delta\tilde U_k(\phi_n)=\tilde U_k(\phi_n)-(-\hf\phi_n^2)$ is plotted
at $\phi_n=n\pi/(10\beta)$ with $n=0,1,\ldots,10$. The gradual
approach of the potential to a parabolic shape is a precursor of the
condensation. According to the numerical results $\Delta\tilde U_k(\phi_n)$
follows a power law behavior in a rather small region above $k_\mr{SI}$.
The exponent characterizing this power law depends on $\beta$, the value $\nu=3.75$
was found at $\beta^2=4\pi$ with $\nu$ decreasing as $\beta$ is increased. 
Once the parabolic shape is approximately installed for 
$k$ slightly above $k_\mr{SI}$ the potential does not suffer a sudden
change with the appearance of the condensate and it assumes its parabolic
shape very rapidly below $k_\mr{SI}$ in the whole period length.

The insertion of an aperiodic parabola section into a periodic potential and the
joining of functions with different analytical structures are 
rather involved issues. The deviation of the dimensionless potential from the parabolic form
around the matching points are found to be weak but not sufficiently fast
converging in the Fourier expansion. The increase of the order of the 
truncation of the Fourier series makes the spread of the parabolic shape
faster at $k\approx k_\mr{IS}$ but the second derivative of the
potential increases at the matching points without bound, too. 
Unfortunately the Wegner-Houghton method does not allow us to go beyond the LPA where 
deviations from the parabolic potential are compatible with the Maxwell-cut.

The dimensionless potential \eq{effpotmol} displays a non-trivial
structure in the molecular phase and is independent of the
bare coupling constants, either relevant or irrelevant at the UV fixed point.
Such a surprisingly strong super-universality renders the deep IR
physics completely parameter-free. Though the UV relevant couplings influence the 
dynamics at finite scales, the physics becomes  parameter-free
well below the condensate scale.

\begin{figure}[ht]
\includegraphics[width=8cm]{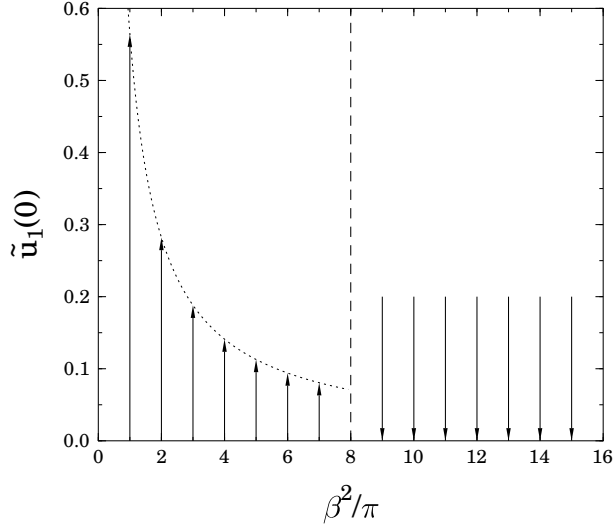}
\caption{
The phase structure of the sine-Gordon model. The dotted line
corresponds to $\tu_1(0) = {\mbox{const.}}/\beta^2$. The critical
value $\beta^2=8\pi$ separates the two phases of the SG model.
\label{sgph}}
\end{figure}

The dimensionless local potential reveals that there are different dynamical mechanisms
leading to the Maxwell-cut and distinguishes the two phases of the SG 
model but this difference is vanishing in fixed, dimensionful units.
Nevertheless the phases can still be distinguished by means of the renormalization group
flow as shown in Fig. \ref{sgph}. It remains to be seen if this difference
rests qualitatively valid when one goes beyond the LPA. In particular,
it would be interesting to decide whether the couplings $\tu_n(k\to 0)$ remain finite 
\cite{Kehrein} or diverge \cite{Amit} beyond the LPA when the parameter 
$\beta$ exhibits also scale-dependence.

{\it Sensitivity matrix} The sensitivity matrix is the systematical tool to distinguish phases of a theory 
according to the global features of the renormalization group flow. Let us 
consider the bare coupling constants, $\tu_n(k)=E_n(\tu(\Lambda),r)$,
as a function of the ratio of the observational scale and the cutoff, $r=k/\Lambda$
and the bare parameters, identified by the initial condition $\tu_n(\Lambda)=E_n(\tu,1)$.
The sensitivity of the renormalization group flow at scale $k$ to an infinitesimal change of
the bare couplings is characterized by the sensitivity matrix \cite{janosRG}
\beq
S_{n,m}(r)=\frac{\partial E_n(\tu(\Lambda),r)}{\partial \tu_m(\Lambda)}.
\eeq
Phase transitions can be identified with the singularities in $S_{n,m}(r)$ as the
function of the parameters of the theory in the limit $r\to0$. The singularities
developing as $r\to0$ with fixed $k$ and $\Lambda\to\infty$ correspond to 
quantum phase transitions because they are driven by short distance phenomena. 
The traditional phase transitions, induced by the low energy, long range modes 
belong to singularities realized with fixed $\Lambda$ and $k\to0$.

\begin{figure}[ht]
\includegraphics[width=8cm]{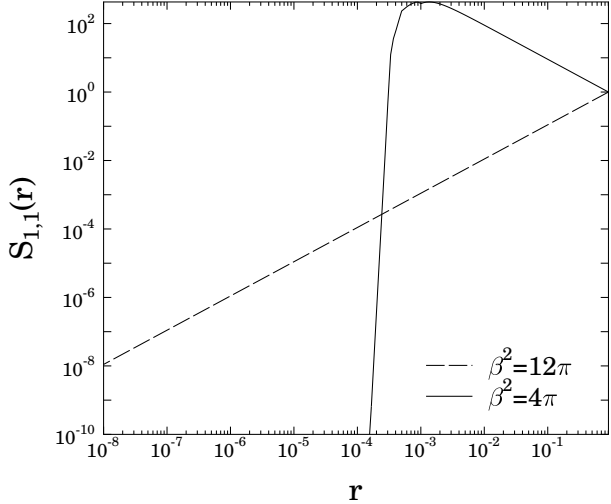}
\caption{The numerically calculated sensitivity matrix at $u_n(\Lambda)=10^{-4}\delta_{n,1}$. 
The slope in the ionized phase is in an excellent agreement with $\eta=1$.
\label{sens}}
\end{figure}

The UV scaling laws are given by Eq. \eq{dluev},
\beq
S_{n,m}(r)= \delta_{n,m} r^{\frac1{4\pi}(n^2\beta^2-8\pi)}
\eeq
in either phase. The situation changes significantly in the IR region, $r\ll 1$, 
where the scaling law
\beq
\tu_n(k)=(-1)^{n+1}R_n \tu_1^n(k)= (-1)^{n+1}R_n\tu_1^n(\Lambda)r^{n\frac1{4\pi}(\beta^2-8\pi)}
\label{smprev}
\eeq
of the ionized phase yields
\beq
S_{n,m}(r)=\delta_{m,1}(-1)^{n+1} R_n
n\tu_1^{n-1}(\Lambda)r^{n\frac1{4\pi}(\beta^2-8\pi)}.
\label{sensmat}
\eeq
The same IR scaling behavior is realized in the molecular phase for $k>k_\mr{SI}$
as in  \eqn{irdlu},
c.f. \fig{sgspin}, thus the \eqn{sensmat} remains valid when $\beta^2<8\pi$.
The low energy dynamics looses sensitivity on any bare coupling constant when
$\eta>0$. This is supported by the numerical results for the sensitivity
matrix, shown in Fig. \ref{sens} with dashed line. The lack of sensitivity is
established in two different ways: both in the UV ($\Lambda\to\infty$) and the
IR ($k\to0$) limits. The renormalized SG model becomes trivial and has no 
free parameter in this phase even if the problem with the linearization of the 
blocking relation around the IR fixed point prevents us to classify the operators 
in the usual fashion.

The sensitivity matrix, given by Eq. \eq{sensmat}, has a singularity as the function of
the coupling constants at $\beta^2=8\pi$ when the limit $r\to0$ is made due to
scaling laws for $k>k_\mr{SI}$, in agreement with the UV origin of the phase transition
in the SG model, c.f. the discussion around \eqn{kc}. The renormalized dynamics at such a scale $k$ has
indeed few free parameters, the relevant coupling constants of the UV
scaling law. This phase transition is shown by the continuous line in 
Fig. \ref{sens} for $r>10^{-3}$. But the approach of the potential to a 
parabolic shape as $k\approx k_\mr{SI}$ stops this trend. The dynamics 
depends less and less  on the high energy, short-distance
parameters as $k$ is decreased below $k_\mr{SI}$ and the super-universality 
developing in this regime renders the dynamics free of any parameter as $k\to0$.

The dimensionful potential shows the same, flat shape in both phases.
Despite the clear difference of the dimensionless potential and the 
sensitivity matrix in the deep IR regime, indicating that the
dynamics, observed in units of the low cutoff differs significantly 
in the two phases, the IR dynamics of these phases do not look different 
when expressed in fixed units. It is easy to understand this apparent
contradiction. The two phases are similar in their IR plane-wave
dynamics, both decouple them. But it is well-known that the SG model
has other important degrees of freedom, solitons, due to the periodicity of the
potential. Their finite size or mass introduces another important scale
where the dynamics differ in the two phases.

{\it Summary} The phase structure and the number of free parameters of the SG
model were studied in this work. The phase structure was identified without
using topological disorder parameters, by means of the renormalization group flow. 
We had to rely on the global analysis of the renormalization group flow due to 
an unexpected problem with the linearization around the IR fixed point of the 
ionized phase. Furthermore, it has been clarified that the sensitivity of 
the dynamics on the renormalizable coupling constants is washed away and 
a super-universality is generated in the IR limit of the molecular phase by the Maxwell-cut.

These features of the SG model show possible interesting analogies with four-dimensional
Yang-Mills theories. Both models have periodic variables. The periodicity
of the gauge model, the fundamental group symmetry is supposed to play
a key role in establishing the confining forces between charges. The renormalizable,
asymptotically free coupling constant of the Yang-Mills theory and the 
SG model in the molecular phase increase with the observational
length scale. Such an increase generates condensate and spinodal
instabilities in the vacuum of both theories. The speed of sound is vanishing
in the mixed phase with spinodal instability, indicating the emergence
of color confinement, the absence of asymptotical charged plane-wave states \cite{gribov}.
The exact scattering matrix constructed in terms of soliton-anti soliton bound states
of the SG model suggests that the asymptotic states of this two dimensional
model are made up by these composite bound states. 
The SG model realizes a mechanism to remove any free parameter in the dynamics
of the condensate, well below its energy scale, in a manner similar to
the long range dynamics of the Yang-Mills vacuum.

\end{document}